\documentclass[aps,prl,preprint]{revtex4-1}
\usepackage{graphicx}
\usepackage{amsfonts}
\usepackage{amsmath}
\usepackage{color}
\usepackage{bm}

\newcommand{\beq}{\begin{equation}}
\newcommand{\eeq}{\end{equation}}

\begin{document}

\title{Elastic and inelastic collisions of $^2\Sigma$ molecules in a magnetic field}
\author{Jie Cui}
\affiliation{Department of Chemistry, University of British Columbia, Vancouver, B.C., V6T 1Z1, Canada}

\author{Roman V. Krems}
\affiliation{Department of Chemistry, University of British Columbia, Vancouver, B.C., V6T 1Z1, Canada}
\date{\today}

\begin{abstract}
We calculate the cross sections for elastic scattering and Zeeman relaxation in binary collisions of molecules  in the ro-vibrational ground state of a $^2\Sigma$ electronic state and the Zeeman state with the electron spin projection $M_S=1/2$ on the magnetic field axis. This is the lowest-energy state of $^2\Sigma$ molecules confined in a magnetic trap. The results are averaged over calculations with multiple molecule - molecule interaction potentials, which yields the expectation intervals for the cross sections and the elastic-to-inelastic cross section ratios. We find that the elastic-to-inelastic cross section ratios under conditions corresponding to trapped molecular ensembles at $T \sim 10^{-3}$ K exceed 100 for the majority of $^2\Sigma$ molecules. The range of $^2\Sigma$ molecules expected to be collisionally unstable in magnetic traps at $T < 10^{-3}$ K is limited to molecules with the spin-rotation interaction constant $\gamma_{\rm SR} > 0.5$ cm$^{-1}$ and the rotational constant $B_e < 4$ cm$^{-1}$.

\end{abstract}

\maketitle

\section{Introduction}

A major focus of current research in molecular physics is the preparation of ultracold molecules \cite{our-book,our-njp-review}. 
While several experimental techniques have been developed for cooling molecules to temperatures of about $10^{-3}$ Kelvin \cite{our-book,our-njp-review,doyle-nature-experiment,stark-deceleration}, it is desirable to obtain dense molecular samples at temperatures below $10^{-6}$ Kelvin \cite{our-njp-review}. 
Molecules at such ultracold temperatures can be produced by photoassociation of ultracold atoms \cite{photoassociation, best-photoassociation-experiments}. However, this technique is limited to molecular species composed of atoms amenable to laser cooling. 
A number of promising applications  -- such as the development of quantum simulators of condensed-matter Hamiltonians \cite{quantum-simulators}, precision measurements of fundamental constants \cite{YbF}, and controlled chemistry \cite{pccp2008} -- need a greater variety of ultracold molecules. Therefore, it is necessary to develop methods for cooling molecules from $10^{-3}$ Kelvin to ultracold temperatures. 
Despite this goal, bridging the milliKelvin -- microKelvin temperature gap remains a major challenge. 

Once cooled to  temperatures $T \sim 1-10$ mK, molecules can be confined in an external field trap. Further cooling of a molecular ensemble can potentially be achieved by evaporation of the most energetic molecules, as is commonly done with alkali metal atoms in magnetic traps \cite{evaporative-cooling}. The evaporative cooling relies on energy thermalization enabled by momentum transfer in molecule - molecule collisions conserving the internal energy of the colliding species. If molecules are trapped in excited states, the evaporative cooling process is hindered by inelastic collisions which release energy and expunge the colliding species from the trap. For example, dc magnetic or electric traps confine molecules in excited Zeeman or Stark states that can undergo collisional energy relaxation \cite{mol-phys-review}, removing particles from the most dense region of the trap. This leads to loss of low-energy molecules and results in heating. The prospect for evaporative cooling of molecules in magnetic traps thus hinges on the relative probability of elastic and inelastic scattering in molecule - molecule collisions in a magnetic field.

The experiments on magnetic trapping of CaH($^2\Sigma^+$) \cite{doyle-nature-experiment}  CrH($^6\Sigma^+$)~\cite{StollPRA08}, MnH($^7\Sigma^+$) ~\cite{StollPRA08} and NH($^3\Sigma^-$)~\cite{CampbellPRL07, TsikataNJP10} in a cold buffer gas of He atoms stimulated multiple theoretical studies of atom - molecule collisions in a magnetic field \cite{bohn, jcp-2004,krems-CaH-2003,Cybulski-Krems-2005, CaF-He-experiment,NH-He-experiment,Hutson-NH-He, Hutson-NH-Mg-1, Hutson-NH-Mg-2, Rb-OH, NH-Rb-Cs, MgH-He,CN-He}. It was found that Zeeman relaxation in atom - molecule collisions is induced by an interplay of intra-molecular spin-dependent interactions and the anisotropy of the  atom - molecule interaction potentials \cite{jcp-2004,krems-CaH-2003,Cybulski-Krems-2005}. A series of experimental \cite{doyle-nature-experiment,CaF-He-experiment,NH-He-experiment} and theoretical \cite{krems-CaH-2003,Cybulski-Krems-2005,Hutson-NH-He} studies showed that collisions of magnetically trapped molecules with weakly interacting He atoms are predominantly elastic, as a consequence of weak interaction anisotropy. At the same time,  it was demonstrated that the rates of Zeeman relaxation in collisions of molecules with more polarizable atoms characterized by a stronger interaction anisotropy are much greater \cite{Hutson-NH-Mg-1, Hutson-NH-Mg-2, Rb-OH,NH-Rb-Cs}. The interaction anisotropy in molecule - molecule collisions is generally much larger than in atom - molecule collisions. 
This raises the question, can molecules be evaporatively cooled? 

The evaporative cooling of molecules in a magnetic trap is certainly possible under specific conditions. Recently, Stuhl et al \cite{stuhl} demonstrated the evaporative cooling of an ensemble of  OH($^2\Pi$) molecules confined in a magnetic trap from 51 to 5.1 milliKelvin. It was shown that inelastic Zeeman relaxation in  OH-OH collisions can be suppressed in this temperature interval due to the peculiar structure of molecules in the $^2\Pi$ electronic state preventing cold molecules from excursions into the strongly interacting regime. The rates of Zeeman relaxation must also be suppressed in the limit of very low temperatures and weak magnetic fields. 
This suppression is a result of the centrifugal barriers which reduce the collision flux into the inelastic scattering channels \cite{bohn, krems-dalgarno-2003}. As a result of this mechanism, molecules cooled to $10^{-6}$ Kelvin or below should be amenable to evaporative cooling in shallow magnetic traps \cite{o2-o2}. However, in order to trap molecules at $10^{-3}$ Kelvin, it is necessary to apply magnetic fields of finite strength ($\gtrsim50$ Gauss). The possibility of evaporative cooling of molecules at such magnetic fields and temperatures remains an open question. 

To answer this question, it is necessary to perform rigorous quantum calculations of molecular scattering properties in the presence of a magnetic field. The theoretical analysis of cold molecular collisions in a magnetic field is, however, complicated by two factors \cite{o2-o2,NH-NH-dipole,NH-NH-magnetic-field,yura-timur-roman}. First, molecule - molecule collisions at low temperatures are affected by scattering resonances, which are sensitive to details of intermolecular interaction potentials. A small variation of the interaction potential may therefore lead to large changes of the collision cross sections.  Second, the interaction of molecules with magnetic fields disturbs the symmetry of the scattering problem. This makes quantum calculations of cross sections for molecule - molecule collisions in a magnetic field prohibitively difficult. If the problem is formulated in the fully uncoupled basis set representation \cite{jcp-2004}, fully converged calculations are impossible \cite{NH-NH-dipole,NH-NH-magnetic-field}. As a result, there are very limited reliable theoretical data for molecule - molecule scattering cross sections in a magnetic field. In particular, nothing is known about the relative probabilities of elastic and inelastic scattering of $^2\Sigma$ radicals, the simplest kind of open-shell molecules with a non-zero magnetic moment.  

  The goal of the present study is to provide predictions of the limits of elastic and inelastic scattering cross sections of $^2\Sigma$ molecules in a magnetic field. To reduce the uncertainty stemming from the limited accuracy of the interaction potentials and scattering resonances, we average the results of quantum scattering calculations over data obtained with multiple interaction surfaces. This produces statistical expectation {\it intervals} of cross sections. We employ the time-independent quantum scattering formalism using the total angular momentum representation that allows large basis sets and dramatically reduces the basis set truncation error \cite{yura-timur-roman}. The calculations are performed as functions of the molecular structure parameters in order to provide useful reference points for future experiments.

\section{Computation details}

We consider collisions of molecules prepared in the ro-vibrational ground state of a $^2\Sigma$ electronic state and the Zeeman state with the electron spin projection $M_S=1/2$ on the magnetic field axis. This is the lowest-energy state of $^2\Sigma$ molecules confined in a magnetic trap \cite{doyle-nature-experiment}.
Evaporative cooling relies on momentum transfer in molecule - molecule collisions that preserve the spin alignment.  
However, molecule - molecule collisions may lead to inelastic relaxation populating the lower-energy Zeeman state characterized by $M_S=-1/2$. Molecules in the state with $M_S=-1/2$ are expunged from the trap by the magnetic field gradient. The ratio of the cross sections for elastic collisions and the Zeeman relaxation $M_S = 1/2 \rightarrow M_S = -1/2$ is thus a critical parameter determining the possibility of evaporative cooling of $^2\Sigma$ molecules in a magnetic trap.

We use the numerical technique developed by Tscherbul and Dalgarno \cite{timur-alex} and Tscherbul, Suleimanov and Krems \cite{yura-timur-roman}. The details of the theory were described previously \cite{jcp-2004,yura-timur-roman,timur-alex}. 
Traditionally, the time-independent quantum calculations of molecular scattering properties were based on representing the total wave function of the collision complex by a basis set expansion in terms of the eigenstates of $\hat{J^2}$ and $\hat{J}_Z$, where  $\hat{J}$ is the total angular momentum of the collision complex and $\hat{J}_Z$ is the $Z$-component of $\hat{J}$ in the space-fixed coordinate frame. The substitution of this expansion in the time-independent Schr\"{o}dinger equation leads to a system of coupled differential equations for the expansion coefficients \cite{arthurs-dalgarno}. In the absence of external fields, the total angular momentum is conserved and the total angular momentum basis leads to a significant reduction of the number of coupled equations. The interaction of molecules with an external field induces couplings between states of different total angular momenta. Therefore, it was originally suggested that the collision theory of molecules in external fields is best formulated using a fully uncoupled space-fixed basis representation \cite{jcp-2004}. The uncoupled basis simplifies the evaluation of the Hamiltonian matrix elements so it was adopted by many authors \cite{He-CH2,uncoupled-o2-o2,uncoupled-N-NH,uncoupled-NH-Li}. 

 If evaluated in the total angular momentum basis, the matrix of the molecule - field interactions is tridiagonal.  
Tscherbul and Dalgarno \cite{timur-alex} and Tscherbul \cite{timur-pra} showed that the tridiagonal character of the Hamiltonian matrix can be exploited for quantum scattering calculations of molecular properties in fields. As described in Refs. \cite{timur-alex, yura-timur-roman, timur-pra}, the scattering cross sections for molecules in external fields can be calculated using the total angular momentum representation with multiple $J$ states included in the basis. The convergence of the scattering cross sections can then be sought with respect to the number of $J$ states. The results of Refs. \cite{timur-alex,yura-timur-roman} show that a few coupled $J$ states are usually sufficient for full convergence of low-energy collision properties in a magnetic field, which makes the total angular momentum basis much more efficient than the fully uncoupled basis. If the total angular momentum basis is used, the collision theory can be 
formulated using either a space-fixed or body-fixed representation \cite{yura-timur-roman}. Here, we use the body-fixed formulation described in detail in Refs.  \cite{timur-alex,yura-timur-roman}. 

We assume that the space-fixed (SF) quantization axis $Z$ is directed along the magnetic field vector and the body-fixed (BF) quantization axis $z$ is directed along the vector $\bm{R}$ joining the centers of mass of molecules $A$ and $B$.
The Hamiltonian for two identical $^2\Sigma$ molecules in a magnetic field can be written (in atomic units) as
\begin{eqnarray}
\hat{H}=-\frac{1}{2\mu R}\frac{\partial^2}{\partial R^2} R + \frac{\hat{l}^2}{2\mu R^2}+\hat{V}(R,\theta_A,\theta_B,\phi)+\hat{V}_{dd}(R,\hat{S}_A,\hat{S}_B)+\hat{H}_{\mathrm{as}}^{(A)}+\hat{H}_{\mathrm{as}}^{(B)}
\label{Hamiltonian}
\end{eqnarray}
where $\mu$ is the reduced mass of the collision complex, 
$
\hat{l}=\hat{J}-\hat{N}_A-\hat{N}_B-\hat{S}_A-\hat{S}_B
$ 
is the orbital angular momentum expressed in terms of the total angular momentum $\hat{J}$, the rotational angular momenta $\hat{N}_A$ and $\hat{N}_B$, and the electronic spin angular momenta $\hat{S}_A$ and $\hat{S}_B$ of the two molecules. The operator $\hat{V}(R,\theta_A,\theta_B,\phi)$ is parametrized by the interaction potential between molecules $A$ and $B$ depending on the polar angles $\theta_i$ of each molecular axis relative to the intermolecular axis ${R}$ as well as the dihedral angle $\phi=\phi_A-\phi_B$.

 The interaction between the unpaired electrons of the molecules gives rise to the intermolecular magnetic dipole - dipole interaction 
\begin{equation}
\hat{V}_\text{dd}(R, \hat{S}_A, \hat{S}_B) = - g_s^2 \mu_0^2 \left( \frac{24\pi}{5}\right)^{1/2} \frac{\alpha^2}{R^3} \sum_q (-)^q Y_{2,-q} (\hat{R}) [\hat{S}_A \otimes \hat{S}_B]^{(2)}_q,
\label{magnetic}
\end{equation}
where $g_s$ is the electron $g$-factor, $\alpha $ is the fine-structure constant, $\mu_0$ is the Bohr magneton and $[\hat{S}_A \otimes \hat{S}_B]^{(2)}_q$ is the tensor product of rank-1 tensors $\hat{S}_A$ and $\hat{S}_B$ \cite{Zare}.
The asymptotic Hamiltonian $\hat{H}_{\mathrm{as}}^{(i)}$ of each $^2\Sigma$ molecule in a magnetic field is given by
\begin{eqnarray}
\hat{H}^{(i)}_{\rm as}=B_e\hat{N}_i^2+\gamma_{\rm{SR}}\hat{N}_i\cdot\hat{S}_{i}+g_s\mu_0B\hat{S}_{Z_{i}}
\end{eqnarray}
where $B_e$ is the rotational constant, $B$ is the magnitude of the magnetic field, $\hat{S}_{Z_{i}}$ is the $Z$-component of $\hat{S}_i$, and $\gamma_{\rm{SR}}$ is the spin-rotation interaction constant.

The eigenstates of the total Hamiltonian are expanded as follows \cite{yura-timur-roman}
\begin{eqnarray}
|\Psi \rangle=\frac{1}{R}\sum_{\alpha_A,\alpha_B}\sum_{J,\Omega}F_{\alpha_A \alpha_B J \Omega}^{M}(R)|N_AK_{N_A}\rangle|S_A\Sigma_A\rangle|N_BK_{N_B}\rangle|S_B\Sigma_B\rangle|JM\Omega\rangle
\label{expansion}
\end{eqnarray}
where $\alpha_i$ denotes collectively the quantum numbers $\{N_i,K_{N_i},S_i,\Sigma_i\}$ of molecule $i$, $K_{N_i}$,$\Sigma_i$ and $\Omega$ are the projections of $\hat{N}_i$, $\hat{S}_i$ and $\hat{J}$ on the BF quantization axis $z$ and $M$ is the projection of $\hat{J}$ on the SF quantization axis $Z$. In the presence of external fields, only $M$ is a good quantum number. This basis set is properly symmetrized to account for the exchange symmetry of identical molecules, as described in Ref. \cite{yura-timur-roman}.

The substitution of the symmetrized expansion into the Schr\"{o}dinger equation yields a set of coupled differential equations. For all computations discussed in the next section, we  integrate these equations on a grid of $R$ from 4.5 ${\rm \AA{}}$ to 500 ${\rm \AA{}}$ using the log-derivative algorithm \cite{johnson1,johnson2} with a step size of 0.05 ${\rm \AA{}}$.  The results presented in Figures 2 and 3 are obtained with the grid of $R$ from 4.5 ${\rm \AA{}}$ to 100 ${\rm \AA{}}$. 
The numerical solutions yield the log-derivative matrix that is converted to the scattering $S$-matrix and the matrix of cross sections using standard equations \cite{jcp-2004}. We perform the calculation with the interaction potential operator defined as $\hat{V} = \lambda V_{\rm NH - NH}$, where $V_{\rm NH - NH}$ is the 4D interaction potential surface for the maximum spin state of the NH - NH dimer recently computed by Janssen {\it et al.} \cite{NH-NH-magnetic-field,NH-NH-free-field}. Each result is averaged over multiple calculations with different $\lambda$ in the interval $\lambda = 0.5-4$.
This interval of $\lambda$ values generates a wide range of interaction potentials. 
When $\lambda = 2.0$, the interaction potential depth is similar to than of the RbCs - RbCs interaction potential \cite{our-2008-paper}, which is dramatically different from that of the NH dimer. The cross sections computed with any given value of $\lambda$ may be affected by resonances and cannot be regarded as meaningful. However, multiple calculations with different values of $\lambda$ provide the reliable expectation intervals of the cross sections. The results of the following sections are presented with one-$\sigma$ standard deviations representing $\sim$68\% confidence intervals.

In order to select the proper basis set, we computed the cross sections for  elastic and inelastic collisions of molecules 
with $\gamma_{\rm{SR}}=0.0415~\rm{cm}^{-1}$ and the rotational constant $B_e=4.2766 ~\rm{cm}^{-1}$. These parameters correspond to the molecule CaH($^2\Sigma^+$). Figure 1 shows the results computed with different basis sets corresponding to the maximum quantum number $N_{\rm max}=2$--$6$ of the rotational angular momentum for each molecule in the expansion (\ref{expansion}). For each calculation, the basis included four total angular momentum blocks corresponding to the fixed total angular momentum projection $M=1$.
Figure 1 illustrates that the elastic scattering cross sections are almost independent of $N_{\rm{max}}$, while the cross sections for the Zeeman relaxation significantly increase with $N_{\rm{max}}$, indicating that it is necessary to include, at least $6$ rotational states ($N=0$--$5$) in the calculations.  

\begin{figure}[!h]
\centering
\includegraphics[width=0.7\textwidth]{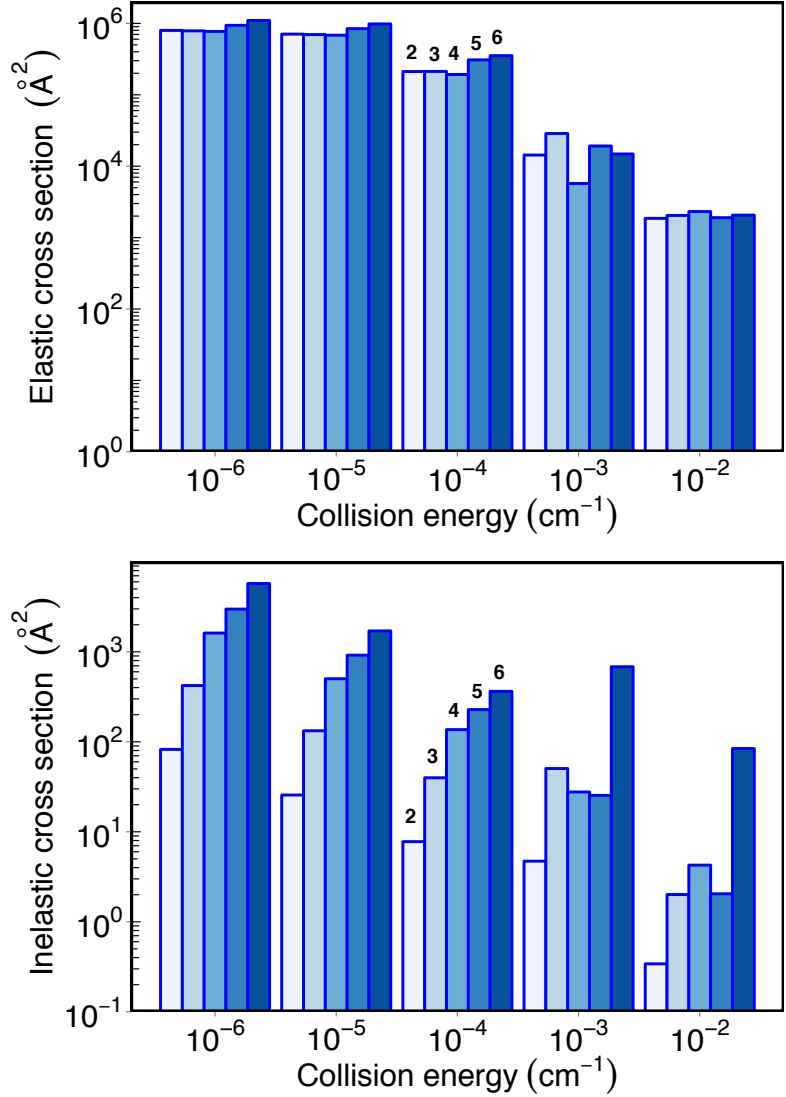}
\caption{(Color online) Cross sections for elastic (upper panel) and inelastic (lower panel) collisions of $^2\Sigma$ molecules with $\gamma_{\rm{SR}}=0.0415  \rm{cm}^{-1}$ and $B_e=4.2766  \rm{cm}^{-1}$ in a magnetic field of 100 Gauss. The colored bars correspond to the different number of rotational states included for each molecule in the basis set expansion (\ref{expansion}), ranging (left to right) from $3$ to $7$.}
\end{figure}

As explained above, elastic and inelastic processes at the collision energies  $\sim 10^{-3}$ K are particularly important for evaporative cooling. The lower panel of Figure 1 shows that the inelastic cross sections at $E= 10^{-3}$ cm$^{-1}$ $\simeq 1.4$ mK are very sensitive to the basis set, exhibiting a significant increase when the maximum number of rotational states in the basis is increased from $6$ to $7$. We have confirmed by the analysis of the energy dependence of the cross sections that this difference is due to a scattering resonance. The results of Figure 1 are obtained with a single potential energy surface corresponding to $\lambda=1$. 
By contrast, Figure 2 displays the elastic and inelastic scattering cross sections averaged over the results of 35 calculations with different $\lambda$ drawn randomly from a uniform distribution in the interval $\lambda = 0.5-4$.

Figure 2 demonstrates that the averaging over the data computed with different interaction potentials reduces the basis set truncation error. To illustrate this point better, we repeated the calculation with the basis set that included the states with $N=0,1$ and $2$ for molecule $A$ and a variable number of rotational states for molecule $B$. This allowed us to examine the role of high-energy rotational states with the rotational angular momentum up to $N=11$. The calculations with a single potential energy surface showed significant sensitivity to the presence of the rotational states with $N<12$. When averaged over the results of 50 calculations with different surfaces, the cross sections appear to be well converged when $N \geq 5$, see Figure 3. The calculations presented in the next section are performed with six rotational states $N=0-5$ for each molecule in the basis set expansion (\ref{expansion}).

\begin{figure}[!h]
\centering
\includegraphics[width=0.6\textwidth]{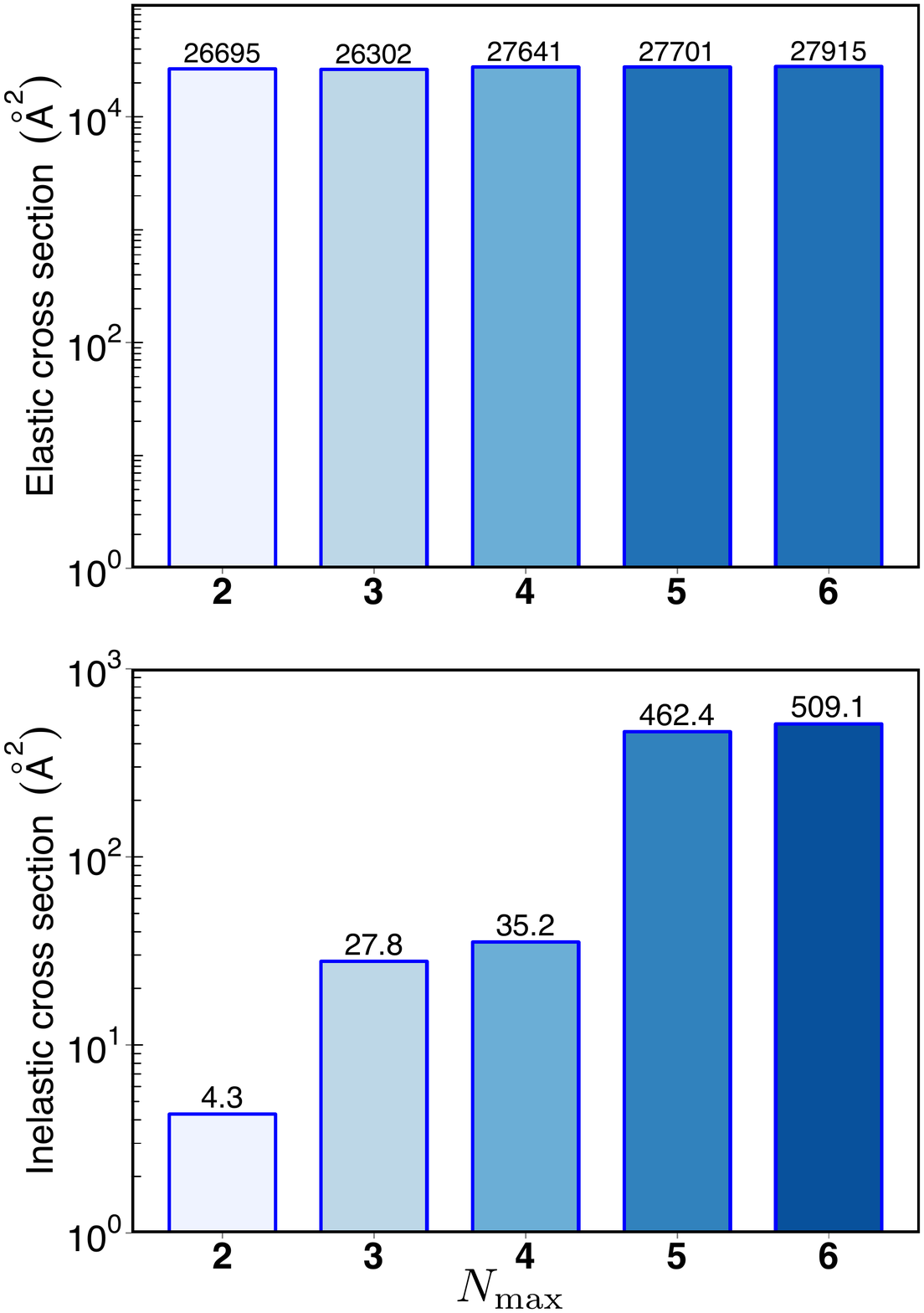}
\caption{(Color online) Cross sections for elastic (left panel) and inelastic (right panel) collisions of $^2\Sigma$ molecules with $\gamma_{\rm{SR}}=0.0415  \rm{cm}^{-1}$ and $B_e=4.2766  \rm{cm}^{-1}$ computed for the collision energy $10^{-3}$ cm$^{-1}$ and the magnetic field 100 Gauss. The colored bars correspond to the different number of rotational states included for each molecule in the basis set expansion (\ref{expansion}), ranging (left to right) from $3$ to $7$.  The results are averaged over 35 calculations with different interaction potentials.}
\end{figure}

\begin{figure}[!h]
\centering
\includegraphics[width=0.6\textwidth]{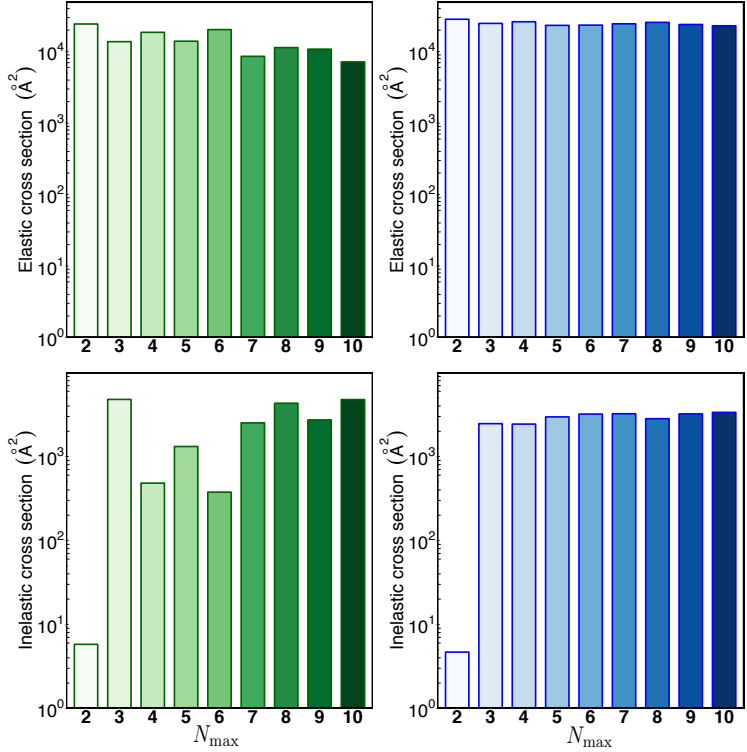}
\caption{(Color online) Cross sections for elastic (upper panels) and inelastic (lower panels) collisions of molecules with $\gamma_{\rm{SR}}=0.0415  \rm{cm}^{-1}$ and $B_e=4.2766  \rm{cm}^{-1}$ computed for the collision energy $10^{-3}$ cm$^{-1}$ and the magnetic field 100 Gauss. The results in the left panels are obtained with a single potential energy surfaces corresponding to $\lambda = 1$. The results in the right panels are averaged over the cross sections computed with 50 different potentials. 
The calculations are performed with $N \leq 2$ for molecule $A$ in the basis set expansion (\ref{expansion}). The colored bars in each panel correspond to the different number of rotational states included for molecule $B$ in the basis set expansion (\ref{expansion}), ranging (left to right) from $3$ to $11$.}
\end{figure}

\section{Spin relaxation mechanisms}

There are two terms in the Hamiltonian (\ref{Hamiltonian}) that induce couplings between the states with $M_S = + 1/2$ and 
$M_S = -1/2$. The magnetic dipole - dipole interaction (\ref{magnetic}) provides direct couplings between the product states $|M_{S_A}=1/2 \rangle|M_{S_B}=1/2 \rangle $ and $|M_{S_A}=-1/2 \rangle|M_{S_B}=-1/2 \rangle$.  In addition, the states with different $M_S$ are coupled by the spin-rotation interaction $\gamma_{\rm SR} \hat{N} \cdot \hat{S}$ in each molecule. 
When both molecules are in the ground rotational state $N=0$, the $M_S =1/2 \leftrightarrow M_S = -1/2$ couplings are induced by a combination of the $N =0 \leftrightarrow N > 0$ couplings due to the anisotropy of the intermolecular interaction potential and the spin-rotation interaction in the states with $N > 0$ \cite{jcp-2004}. The cross sections for inelastic collisions leading to spin
relaxation $M_S = +1/2 \rightarrow M_S = -1/2$ in one or both molecules are determined by the competition of these two mechanisms. While the magnetic dipole - dipole interaction is fixed, the dynamical processes induced by the spin-rotation interaction depend on the magnitude of the spin-rotation interaction constant and the rotational constant of the molecule \cite{jcp-2004}. 

In order to explore the dependence of the cross sections for elastic and inelastic scattering on $\gamma_{\rm SR}$, we average the calculations over the results obtained with 48 interaction potentials. The calculations are performed for molecules with the rotational constant $B_e=4.2766$ cm$^{-1}$, the collision energy $10^{-3}$ cm$^{-1}$ and a magnetic field magnitude of $100$ Gauss.  This particular magnetic field magnitude is chosen because the potential energy shift of $^2\Sigma$ molecules in the $|N=0, M_S=1/2 \rangle$ state due to the Zeeman effect at 100 Gauss is about $4.7 \times {10}^{-3}$ cm$^{-1}$ = $6.7 \times 10^{-3}$ K. A magnetic trap with a field gradient up to 100 Gauss can thus confine molecular ensembles with temperatures $ T < 5\times10^{-3}$ K. In the limit of vanishing magnetic field, the low-energy cross sections for the Zeeman relaxation $M_S=1/2 \rightarrow M_S = -1/2$ tend to zero \cite{krems-CaH-2003}.  As the field increases from zero, the inelastic cross sections in the absence of resonant scattering generally increase and the elastic cross sections are independent of the magnetic field \cite{o2-o2}. The results at $B = 100$ Gauss thus represent the highest probability of inelastic scattering in a magnetic trap with molecules at $T < 5\times 10^{-3}$ K. As the temperature of the molecular ensemble is reduced, the depth of the magnetic trap can be lowered, leading to smaller cross sections for inelastic scattering and higher ratios of elastic-to-inelastic cross section \cite{o2-o2}. 

We note that in addition to spin relaxation, $^2\Sigma$ molecules in a magnetic trap may also undergo chemical reactions through non-adiabatic transitions to the singlet spin state of the collision complex \cite{pccp2008}. These non-adiabatic transitions are induced by the same couplings as the inelastic Zeeman relaxation. Such chemical processes have been explored for the case of $^3\Sigma$ molecules \cite{gerrit}.  Unlike Zeeman relaxation, chemical reactions produce molecular fragments with high kinetic energy. Therefore, the suppression mechanism mentioned above does not apply to chemical reactions and the reaction rates are largely independent of the magnetic field. Because the chemical reactions and the Zeeman relaxation are mediated by the same couplings, the rates of these two processes are very similar at high fields. The calculations in Ref. \cite{gerrit} showed that the rates for the chemical reactions and the spin relaxation in NH - NH collisions are very similar at 100 Gauss. For heavier molecules, the Zeeman relaxation rates must approach the reaction rates at lower magnetic fields.

 Figure 4 shows the averaged results for the elastic and Zeeman relaxation cross sections with the $\pm \sigma$ intervals and illustrates two important observations. 
 First, the elastic-to-inelastic ratio is very large at low magnitudes of $\gamma_{\rm SR}$. This indicates that the magnetic dipole - dipole interaction is relatively ineffective. Molecules with small spin-rotation interaction constants are thus likely to be amenable to evaporative cooling.  This is in agreement with a recent calculation of the Zeeman relaxation  cross-sections in collisions of CaH($^2\Sigma^+$) molecules with Li ($^2S$) atoms \cite{tscherbul-coworkers}. In this paper, Tscherbul and coworkers showed that collision-induced spin-relaxation may be very inefficient even in systems with very large interaction anisotropy, providing it is mediated by weak spin-dependent interactions.  The spin-dependent interactions serve as a bottleneck suppressing the Zeeman relaxation even in the limit of infinitely strong interaction anisotropy \cite{tscherbul-coworkers}.  

Second, Figure 4 shows that the inelastic cross sections can be well approximated by a quadratic function of $\gamma_{\rm SR}$. This indicates that the spin-relaxation process is largely driven by second-order interactions involving the matrix elements of the spin-rotation interaction \cite{jcp-2004}. The role of the magnetic dipole-dipole interaction is apparent at low values of $\gamma_{\rm SR}$, where the dependence of the Zeeman relaxation cross sections on $\gamma_{\rm SR}$ is weak. The strong dependence of the inelastic cross sections on $\gamma_{\rm SR}$ at $\gamma_{\rm SR} > 0.4$ cm$^{-1}$ suggests that the mechanism of the Zeeman relaxation at these strengths of the spin-rotation interactions is dominated by the $\gamma_{\rm SR}$-dependent couplings. The elastic-to-inelastic ratio is inversely proportional to $\gamma_{\rm SR}$ and appears to be above 100, even for molecules with large spin-rotational constants $\sim 1$ cm$^{-1}$ \cite{mizushima-book}.  This value of the elastic-to-inelastic ratio is considered to be critical for the efficacy of evaporative cooling and is often used as a limit below which the cooling experiments become unfeasible \cite{Egorov}.

\begin{figure}[h!]
\centering
\includegraphics[width=0.8\textwidth]{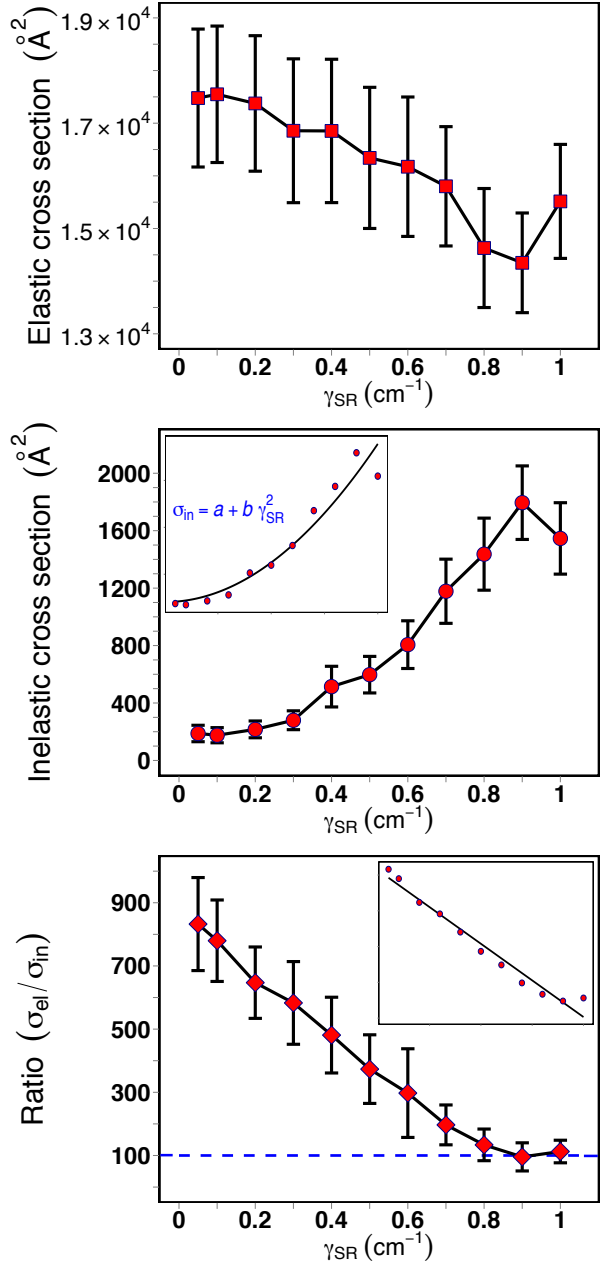}
\caption{(Color online) Cross sections for elastic (upper panel) and inelastic (middle panel) collisions of molecules with  $B_e=4.2766 ~\rm{cm}^{-1}$ computed as functions of $\gamma_{\rm SR}$ for the collision energy $10^{-3}$ cm$^{-1}$ and the magnetic field $100$ Gauss.  The lower panel shows the elastic-to-inelastic  cross section ratio.
The results are averaged over 48 calculations with different interaction potentials. The vertical bars indicate the 2$\sigma$ interval of the cross section values and their ratios. The insets illustrate the polynomial regression fits.}
\end{figure}

In the case of collisions with atoms, the Zeeman relaxation of molecules in the state $|N=0, M_S = 1/2 \rangle$ is very sensitive to the energy separation between the rotational states \cite{CaF-He-experiment, prl-2006,MgH-He,CN-He,LiH-He}. In order to examine the sensitivity of the Zeeman relaxation in molecule - molecule collisions to the rotational energy splittings, we present in Figures 5 and 6 the elastic and inelastic cross sections as functions of the rotational constant. The calculation in Figure 5 is for molecules with a very small value of $\gamma_{\rm SR} = 0.0415$ cm$^{-1}$ and the results in Figure 6 are for molecules with $\gamma_{\rm SR} = 0.5$ cm$^{-1}$. When the spin-rotation interaction is weak, the elastic and inelastic cross sections exhibit a weak dependence on the rotational constant of the molecule. This indicates that, at this magnitude of the spin-rotation interaction, the Zeeman relaxation is largely driven by the magnetic dipole - dipole interaction. By contrast, the inelastic cross sections displayed in Figure 6 decrease monotonically as the rotational constant $B_e$ increases.  This is another evidence that the mechanism of the Zeeman relaxation induced by the spin-rotation interaction clearly dominates at these values of $\gamma_{\rm SR}$. The results presented in Figure 6 show that the elastic-to-inelastic ratios may exceed 100 even for molecules with the spin-rotation interaction constant as large as $\gamma_{\rm SR} = 0.5$ cm$^{-1}$, providing the rotational constant is greater than $4$ cm$^{-1}$. The range of $^2\Sigma$ molecules expected to be collisionally unstable in magnetic traps at $T < 10^{-3}$ K is thus restricted to molecules with $\gamma_{\rm SR} > 0.5$ cm$^{-1}$ and $B_e < 4$ cm$^{-1}$. 

\begin{figure}[h!]
\centering
\includegraphics[width=0.8\textwidth]{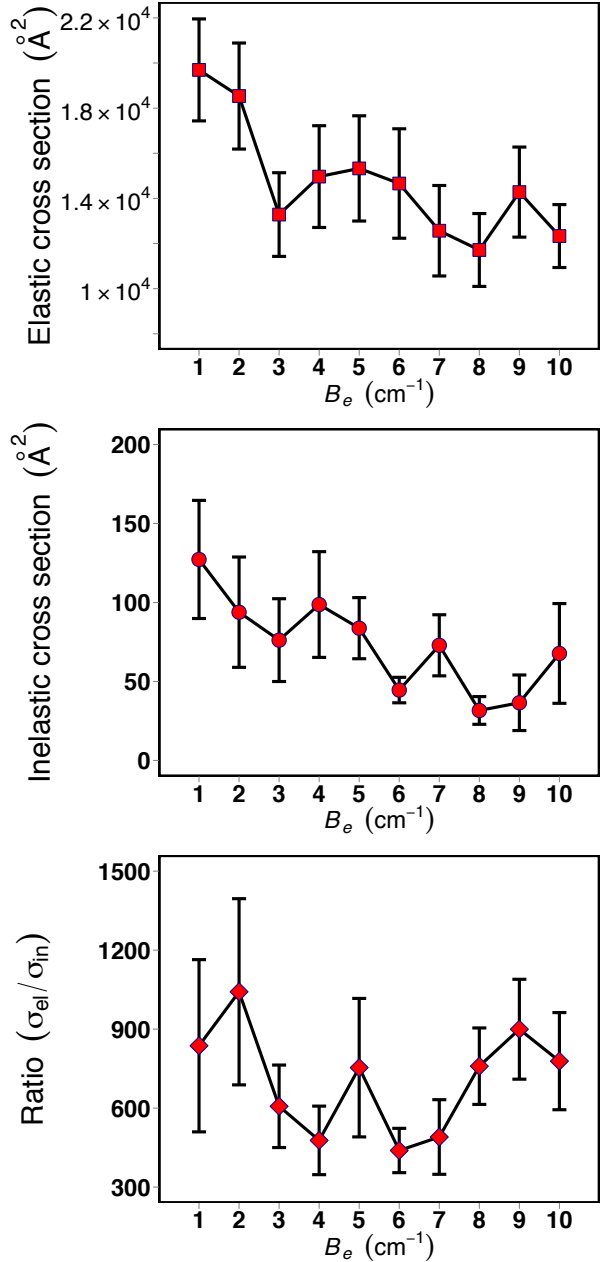}
\caption{(Color online) Cross sections for elastic (upper panel) and inelastic (middle panel) collisions of molecules with $\gamma_{\rm SR}=0.0415$ cm$^{-1}$ computed as a function of $B_e$ for the collision energy $10^{-3}$ cm$^{-1}$ and the magnetic field $100$ Gauss. The lower panel shows the elastic-to-inelastic  cross section ratio. The results are averaged over 20 calculations with different interaction potentials. The vertical bars indicate the $2\sigma$ interval of the cross section values and their ratios. }

\end{figure}

\begin{figure}[h!]
\centering
\includegraphics[width=0.8\textwidth]{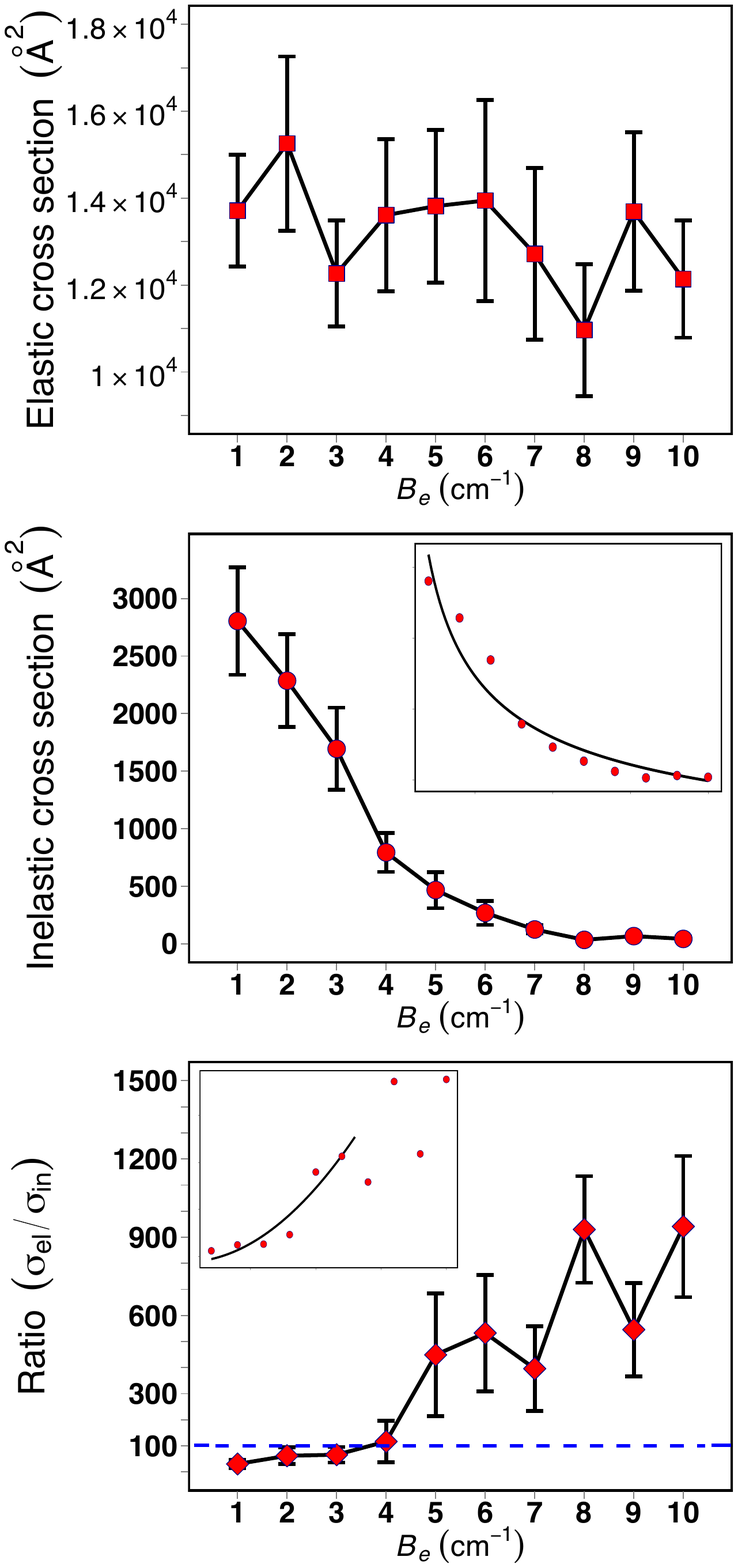}
\caption{(Color online) Cross sections for elastic (upper panel) and inelastic (middle panel) collisions of molecules with $\gamma_{\rm SR}=0.5$ cm$^{-1}$ computed as a function of $B_e$ for the collision energy $10^{-3}$ cm$^{-1}$ and the magnetic field $100$ Gauss. The lower panel shows the elastic-to-inelastic  cross section ratio. The results are averaged over 35 calculations with different interaction potentials. The vertical bars indicate the 2$\sigma$ interval of the cross section values and their ratios. The insets illustrate the polynomial regression fits.}
\end{figure}

\section{Conclusion}

Diatomic molecules in the $^2\Sigma$ electronic state represent the simplest class of molecular radicals that can be confined in a magnetic trap. While the magnetic trapping of $^2\Sigma$ molecules has been achieved, the feasibility of evaporative cooling of $^2\Sigma$ molecules to ultracold temperatures depends on the magnitudes of the cross sections for elastic scattering and inelastic Zeeman relaxation  and is yet to be experimentally demonstrated. The present work provides insight into the relative magnitudes of these cross sections. In order to draw general conclusions, we performed the computations with multiple potential energy surfaces, yielding the expectation intervals for both the elastic and inelastic scattering cross sections. 

  The first important result of this work is illustrated by Figures 1, 2 and 3. These calculations show that the cross sections averaged over multiple potential energy surfaces are much less sensitive to the basis set truncation error. This indicates that statistically meaningful results can be obtained by averaging over multiple calculations with fairly restricted basis sets. This also made the calculations presented in Figures 4, 5 and 6 feasible. Reducing the number of rotational states in the basis set from $7$ to $6$ reduces the computation time of each data point on a fast CPU from about 1 month to about one week. It thus took about 48 weeks of CPU time to compute each interval depicted in Figure 4. Averaging over multiple interaction potentials also allows for the analysis of the expectation values of the cross sections as functions of the molecular parameters.

We analyzed the results as functions of the molecular parameters (the spin-rotation interaction strength and the rotational constant magnitude) for the collision energy $10^{-3}$ cm$^{-1}$ and the magnetic field $100$ Gauss. This collision energy corresponds approximately to the energy of molecules at about $1.4\times 10^{-3}$ K. This magnetic field magnitude is representative of the field strength experienced by molecules in magnetic traps with the trap depth $\sim 5 \times 10^{-3}$ K. The inelastic cross sections are expected to be smaller and the elastic-to-inelastic ratios higher at lower magnetic fields. 

Our results show that the elastic-to-inelastic ratios are consistently greater than 100, except for molecules with very large constants of the spin-rotation interaction and small rotational constants ($B_e < 4$) cm$^{-1}$. This 
indicates a good prospect for evaporative cooling of a great variety of $^2\Sigma$ molecules to temperatures below $10^{-3}$ K. Molecules with small constants of the spin-rotation interaction appear to be the best candidates for evaporative cooling, irrespective of the magnitude of the rotational constant. 
This challenges the previous conclusion \cite{jcp-2004} that only molecules with large rotational constants should be amenable to collisional cooling. Molecules with the spin-rotation interaction constants $> 0.4$ cm$^{-1}$ have favourable collision properties only if their rotational constant $B_e > 5$ cm$^{-1}$. Refs. \cite{mizushima-book} and \cite{Egorov} provide a list of the spectroscopic constants for a variety of diatomic molecules in the $^2\Sigma$ electronic state. Figure 7 displays the lower limit of the expectation values of the elastic-to-inelastic ratios for these molecules obtained by the interpolation of our results.
To obtain these data, we used the lowest point of a $2\sigma$ deviation of the elastic-to-inelastic ratio from the mean value, corresponding to the $\sim 95$~\% confidence interval. 
 The figure illustrates that the majority of the selected $^2\Sigma$ radicals should be amenable to evaporative cooling starting at milliKelvin temperatures. It is important to emphasize that the results presented in Figure 7 should be interpreted as expectation values accurate to within $\sim$95~\%. In any specific case, the collisional interaction between the molecules may be affected by details of the interaction potentials, leading to lower elastic-to-inelastic cross section ratios. Nevertheless, the results of Figure 7 are very encouraging for the prospect of evaporative cooling of $^2\Sigma$ molecules as most of the data appear to be well above 100.

\begin{figure}[h!]
\centering
\includegraphics[width=0.6\textwidth]{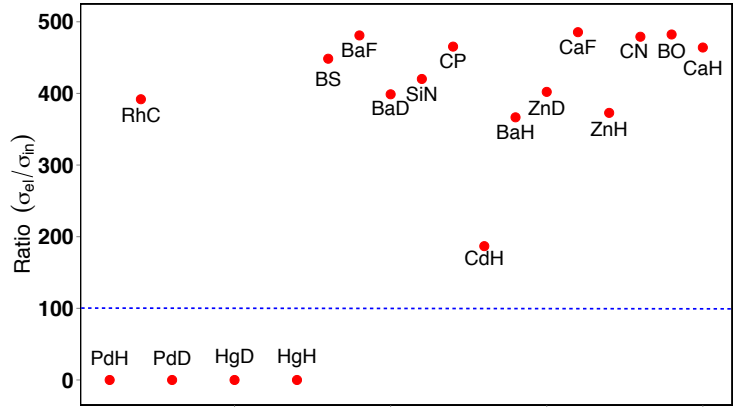}
\caption{(Color online) The lowest magnitude of the expected elastic-to-inelastic cross section ratio for selected $^2\Sigma$ molecules. The results are obtained by interpolation of the data in Figures 4, 5 and 6 using the spectroscopic constants borrowed from Ref. \cite{mizushima-book}.}
\end{figure}

\section{Acknowledgment}

This work is supported by NSERC of Canada.

\end{document}